\documentclass[a4paper,11pt]{article}
\begin{document}

\title{A proposed observation during the \\total solar eclipse on 11
August 1999}

\author{{\bf Amitabha Ghosh} \\Director, Indian Institute of
Technology\\
Kharagpur--721302, India.\\ E-mail: director@iitkgp.ernet.in\\ and \\
{\bf Soumitro
Banerjee}\\Department of
Electrical Engineering\\ Indian Institute of Technology\\
Kharagpur--721302,
India.\\ E-mail: soumitro@ee.iitkgp.ernet.in}  
\date{} \maketitle  

\abstract{The theory of inertial induction, which has been quite
successful in explaining a number of hitherto unexplained astrophysical
phenomena, predicts a redshift in light grazing massive objects. The total
solar eclipse on 11 August 1999 offers an opportunity of testing the
theory. In this paper we predict the additional redshift of a few stars
during the eclipse and urge professional astronomers to do the
observation.}

\bigskip \bigskip

\noindent {\bf Keywords:} Inertial induction, total solar eclipse,
redshift.

\newpage

The number of astronomical phenomena that cannot be explained in the light
of conventional physics has built up over the years. The discrepency
between the astrophysical mass and relativistic mass of white dwarfs has
been a long-standing problem [\ref{white}]. It has been reported that the
spectrum of the light coming from the edge of the sun shows a redshift
different from that of the radiation coming from the center [\ref{AG2}].
It is known that the Earth's spin is slowing down at the rate of $6\times
10^{-22} \mbox{rad s}^{-2}$. If the tidal friction theory is
applied to explain this, one encounters moon's close approach problem
[\ref{AG2}]. It has been found that Phobos experiences a secular
acceleration of $1.05\times 10^{-20} \mbox{rad s}^{-2}$ and Deimos
experiences a secular retardation of the order of $2.46\times 10^{-23}
\mbox{rad s}^{-2}$ [\ref{phobos}]. In addition, it has been reported that
radiation grazing past the sun experiences an anomalous redshift. Sadeh
et. al. [\ref{sadeh}] reported that the 21 cm signal coming from Taurus~A
suffered a redshift of 150 Hz while grazing the sun at a distance of 5
solar radii on 15 June 1967. Merat et. al. [\ref{merat}] reported that the
2292 MHz signal from Pioneer-6 was also found to be subjected to a
redshift when it passed behind the sun. There is no satisfactory
quantitative explanation of these phenomena.

This has led many physicists to believe that the theory of gravitation
needs to be modified to account for the observed facts, and one of the 
emerging theories is that of inertial induction. The theory of
inertial induction proposes that gravitational interaction between two
bodies depends not only on the relative separation between the two bodies,
but also on the relative velocity and acceleration. Many functional forms
of the inertial induction terms have been proposed, out of which the
following form [\ref{AG2},\ref{AG1}] seems to account for the maximum
number of astrophysical phenomena:

\begin{equation} {\bf F} = - \frac{Gm_Am_B}{r^2}\hat{\bf u}_r -
\frac{Gm_Am_B}{c^2r^2} v^2f(\theta )\hat{\bf u}_r - \frac{Gm_Am_B}{c^2r} a
f(\phi )\hat{\bf u}_r \label{gravity} \end{equation} where $m_A$ and $m_B$
are the gravitational masses of $A$ and $B$ respectively, $v$ and $a$ are
the magnitudes of the relative velocity and acceleration of $A$ with
respect to $B$, $\hat{\bf u}_r$ is the unit vector along ${\bf r},
f(\theta)$ and $f(\phi)$ (with $\cos\theta = \hat{\bf u}_r. \hat{\bf
u}_v$
and $\cos\phi = \hat{\bf u}_r . \hat{\bf u}_a)$ represent the inclination
effects. $f(\theta)$ is assumed to be symmetric, satisfying the conditions
(a) $f(\theta)\!=\!1$ for ${\theta}\!=\!0$, (b) $f(\theta)\!=\!-1$ for
$\theta\!=\!\pi$, and (c) $f(\theta )\!=\!0$ for $\theta\!=\!  \pi/2$.
$f(\phi)$ is assumed to have the same functional form as $f(\theta)$. The
first term in (\ref{gravity})  represents the ``static'' interaction of
Newtonian gravitation, the second term represents a dynamic velocity
dependent inertial induction and the third term represents an acceleration
dependent inertial induction. This theory postulates that gravity can act
on all forms of matter and radiation --- in the latter case one has to
take the relativistic mass $E/c^2$.

The theory of inertial induction as quantified by equation (\ref{gravity})
leads to a number of very intriguing consequences. Some of these are
enumerated below (detailed derivation can be found in [\ref{AG2}]).

\begin{enumerate}

\item The model leads to the exact equivalence of gravitational and
inertial masses, and the derivation of a new force law 
\[{\bf F}=m{\bf a}+\mbox{a very
small cosmic drag term.}\]

\item The model provides a mechanism for the transfer of angular momentum
from a central body to the orbiting objects and explains the transfer of
solar angular momentum.

\item The model predicts slowing down of the Earth's spin at the rate of
$5.5\times 10^{-22} \mbox{rad s}^{-2}$.  It predicts a secular
acceleration
of Phobos to the extent of $0.81\times 10^{-20} \mbox{rad s}^{-2}$. In
case of Deimos the predicted deceleration is $4.94\times 10^{-23}
\mbox{rad s}^{-2}$. The predictions match the observed values remarkably
well (in case of Deimos the observation is not accurate and the standard
error of measurement is 3 times the mean value).

\item The model predicts excess redshift of light in the presence of
massive objects. This resolves the long-standing discrepency between
the astrophysical mass and relativistic mass of white dwarfs. It
quantitatively explains the excess redshift in the spectrum of the solar
limb. It also predicts excess redshift of radiation grazing massive
objects, thus explaining the observed redshift of the radiation
from Taurus~A and Pioneer-6 at near occultation position with respect to
the sun.

\end{enumerate}

In spite of the above successes, we believe that the strength of a good
theory lies in its ability to predict the outcome of experiments {\em
before} these are performed --- which constitutes its falsifiability
criterion. In this paper we lay down two clear predictions which can be
tested in near future.

First, the theory predicts a slowing down of Mars' spin at the rate of
$1.25\times 10^{-22} \mbox{rad s}^{-2}$, but no experiment/observation has
yet been attempted to detect this. We suggest that efforts should be
directed towards this observation. If such a secular retardation of Mars
is detected, in the absence of tidal effects there will be no other
explanation.

Second, the theory predicts a redshift in starlight grazing the sun --- an
observation which can be made during a total solar eclipse. The magnitude
of the redshift is 
\begin{equation}
z=\frac{\Delta
\lambda}{\lambda}=\exp\left(\frac{4GM_\odot}{3c^2d}\right)-1
\approx \frac{4GM_\odot}{3c^2d} \label{z}
\end{equation}
where $d$ is the minimum distance of the path of the starlight from the
sun.

 The next total solar eclipse will take place on 11th August 1999.
Unfortunately there will not be any bright star near the sun at the time
of the eclipse, but we suppose the modern gadgets can detect relatively
faint ones --- possibly at other wavelengths. In Table~1, we give the
predicted values of the incremental redshift for a few stars which will be
close to the solar disk at the time of the eclipse. The incremental
redshifts of other stars near the sun can be predicted using Eqn.~\ref{z}.

\noindent {\bf Acknowledgments:} We thank Mr. Subhasis Basak of the Saha
Institute of Nuclear Physics, Calcutta, India, for his help in
identifying the stars which will be close to the sun at the time of the
eclipse.

\small
\begin{table}[t]
\caption{The predicted incremental redshift of a few stars during the
total solar eclipse 11 August 1999. We have assumed that the sun's
position during the eclipse will be  9h23m30s R.A. and
$15^o 17^\prime 0^{\prime\prime}$ Dec.} 
\vspace*{0.2in}
\begin{tabular}{llllll} \hline
Star & Const. & Mag. & R.A. & Dec. & Predicted  $z$\\
\hline
82 Cnc & Cancer & 5.4 &  9h 15m 13.8s &  $14^o 56^\prime
29.4^{\prime\prime}$  &
$3.7\times
10^{-7}$\\
XZ14280 & 12BLeo & 6.3 & 9h 25m 32.5s & $16^o 35^\prime
08.1^{\prime\prime}$  & $5.6\times
10^{-7}$\\                                                    
XZ14313 & Leo & 6.8 & 9h 26m 56.7s & $14^o 18^\prime 11.2^{\prime\prime}$  
& $6.0\times
10^{-7}$\\                                                         
XZ14117 & Leo & 6.6 & 9h 18m 58.8s & $17^o 04^\prime 19.3^{\prime\prime}$  
& $3.7\times
10^{-7}$\\                                                 
XZ14433 & 5Leo/xi-Leo & 5.0 & 9h 31m 56.7s & $11^o 17^\prime
59.4^{\prime\prime}$  &
$1.7\times
10^{-7}$\\                                                         
XZ14434 & 6Leo/h-Leo & 5.1 & 9h 31m 57.6s & $09^o 42^\prime
56.8^{\prime\prime}$  & $1.3
\times
10^{-7}$\\                                                         
\hline
\end{tabular}
\end{table}
\normalsize

\subsection*{References:}

\begin{enumerate}

\item Greenstein, J. L. and Trimble, V. L., {\em Astrophysical Journal},
{\bf 149}, 283 (1967); Moffett T. J. et. al., {\em Astronomical Journal},
{\bf 83}, 820 (1978); Shipman, H. L. and Sass, C. A., {\em Astrophysical
Journal}, {\bf 235}, 177 (1980); Grabowski, B. et. al. {\em Astrophysical
Journal}, {\bf 313}, 750 (1987). \label{white} 

\item Ghosh, A. in {\em ``Progress in New Cosmologies: Beyond the Big
Bang},'' H. C. Arp et. al. (Ed.), Plenum Press, (1993). \label{AG2}

\item Sinclair, A. T., {\em Astronomy and Astrophysics}, {\bf 220}, 321
(1989). \label{phobos}

\item Sadeh, D., Knowles, S. H. and Yaplee, B. S., {\em Science}, {\bf
159}, 307 (1968). \label{sadeh}

\item Merat, P., Pecker, J-C. and Vigier, J-P., {\em Astronomy and
Astrophysics}, {\bf 174}, 168
(1974). \label{merat}

\item Ghosh, A., {\em Pramana (Journal of Physics)}, {\bf 26}, 1 (1986).
\label{AG1}

\end{enumerate}

\end{document}